\documentstyle[psfig,PASJadd]{PASJ95}
\newcommand{\vol}{}
\newcommand{\no}{}
\newcommand{\lett}{}
\newcommand{\spage}{}
\newcommand{\rdate}{}
\newcommand{\adate}{}
\begin{document}
\title{A New Method to Estimate Cosmological Parameters Using
Baryon Fraction of Clusters of Galaxies}
\author{Shin {\sc Sasaki} \\
{\it Department of Physics, Tokyo Metropolitan University, Hachioji,
Tokyo, 192-03} \\
{\it E-mail: sasaki@phys.metro-u.ac.jp}}
\abst{We propose a new method to estimate cosmological parameters using the
baryon fraction of clusters of galaxies for a range of redshifts.
The basic assumption is that the baryon fraction of clusters is
constant, which is a reasonable assumption when it is averaged
within a Mpc scale.
We find that the baryon fraction vs. redshift diagram can
estimate the  cosmological parameters, since the derived value of the
baryon fraction from observations depends on the adopted value of
cosmological parameters. 
We also discuss some points in comparing theoretical calculations
with observations.}
\kword{Cosmology --- Galaxies : clusters of --- X-rays}
\maketitle
\thispagestyle{headings}

\section{Introduction}

Clusters of galaxies are the largest virialized systems in the
universe and their masses can be estimated by X-ray and
optical observations.
In general, since there is no efficient way of changing the baryon
fraction averaged within a Mpc scale, the baryon fraction of a cluster 
of galaxies $M_b/M_{tot}$ ($M_b$ and $M_{tot}$ are baryon and
total masses of a cluster) should take the global fraction in the 
universe: $ \Omega_b/\Omega_0 $ 
($\Omega_b$ and $\Omega_0$ are the baryon density parameter and
the total density parameter).
White et al. (1993) estimated $\Omega_0$ by identifying $ M_b /
M_{tot} $ of the Coma cluster with $ \Omega_b / \Omega_0$ and by using
the big bang nucleosynthesis value for $ \Omega_b$, 
and they found that low $\Omega_0$ value is favored:
$ \Omega_0 = 0.16 h^{-1/2} / (1 + 0.19 h^{3/2}) $  (where $H_0 = 100 h 
{\rm ~km~s^{-1}~Mpc^{-1}} $ is the Hubble constant).
However, in this method, the derived $ \Omega_0 $ depends on the
Hubble constant and studies of the big bang nucleosynthesis. 
Furthermore, this method does not probe the value of the cosmological
constant $\lambda_0$.

In this {\it Letter}, we propose a new method to estimate cosmological 
parameters (both $\Omega_0$ and $\lambda_0$) using the baryon fraction 
of clusters for a range of redshifts.

\section{Cosmological parameter dependence on mass estimation}

First of all, we briefly summarize the method to estimate gas,
galaxy and total masses of a cluster from X-ray and optical
observations paying attention to their cosmological parameter
dependence.


Firstly, we describe the gas mass in a cluster of galaxies using X-ray 
data.
For simplicity, we assume that the hot gas in a cluster follows a
spherically symmetric isothermal $\beta -$ model (e.g., Sarazin 1988): 

\begin{equation}
n_e (r) = n_{e0} \left( 1 + \frac{r^2}{{r_c}^2} \right)^{-3 \beta/2}, 
\end{equation}
where $n_e$ is the electron number density, and $ n_{e0}$ and $ r_c$
are the central electron density and the core radius, respectively.
Then, using X-ray data, the gas mass $ M_{gas} (<R) $ within a radius
$R$ and the bolometric luminosity $ L_X(<R)$ derived by X-ray
observation are written as 

\begin{eqnarray}
M_{gas} (<R) &=& \int_0^R \rho_{gas}(r) 4 \pi r^2 dr \nonumber\\
&=& \frac{8 \pi}{1+X} m_H n_{e0} {r_c}^3 I_M (R/r_c,\beta), \\
I_M (y, \beta) &\equiv& \int_0^y (1+x^2)^{-3 \beta/2} x^2 dx, \\
L_X (<R) &=& \int_0^R \left(\frac{2 \pi k_B T_e}{3 m_e} \right)^{1/2}
\left( \frac{2^4 e^6}{3 \hbar m_e c^2} \right) \nonumber\\
& & \times [n_e(r)]^2
\bar{g_B}(T_e) \frac{2}{1+X} 4 \pi r^2 dr\nonumber\\
&=&  \left(\frac{2 \pi k_B T_e}{3 m_e} \right)^{1/2}
\left( \frac{2^4 e^6}{3 \hbar m_e c^2} \right) \bar{g_B}(T_e) \nonumber\\
& & \times
\frac{2}{1+X} 4 \pi {n_{e0}}^2 {r_c}^3 I_L(R/r_c,\beta), \\
I_L (y, \beta) &\equiv& \int_0^y (1+x^2)^{-3 \beta} x^2 dx,
\end{eqnarray}
where $\rho_{gas}$ is the gas density and other symbols have their
usual meanings. 
Combining equations (2) and (4) yields the gas mass:

\begin{eqnarray}
M_{gas} (<R) &=& \left( \frac{3 \pi \hbar m_e c^2}{2 (1+X) e^6}
\right)^{1/2}  \left( \frac{3 m_e c^2}{2 \pi k_B T_e} \right)^{1/4}
m_H \nonumber\\
& & \mbox{\hspace{-2.5cm}} \times \frac{1}{[\overline{g_B}(T_e)]^{1/2}} 
{r_c}^{3/2} \left 
[ \frac{I_M (R/r_c, \beta)}{I_L^{1/2} (R/r_c, \beta)} \right] [L_X
(<R)]^{1/2}.
\end{eqnarray}
In the above equations, we implicitly assumed that the X-rays are emitted
by the thermal bremsstrahlung only (see e.g., Rybicki and Lightman 1979).
Note that $L_X, r_c$ and $R$ are not the quantities derived by
observations directly, but that they depend on the adopted cosmological
parameters.
That is, they are estimated as

\begin{eqnarray}
L_X (<R) &=& 4 \pi [D_L(z; \Omega_0, \lambda_0, H_0)]^2 f_X(<\theta), 
\\
r_c &=& \theta_c D_A(z; \Omega_0, \lambda_0, H_0), \\
R &=& \theta D_A(z; \Omega_0, \lambda_0, H_0),
\end{eqnarray}
using directly observed quantities: the total bolometric flux $
f_X(<\theta)$ 
within $\theta$, the angular core radius
$\theta_c$ and the outer angular radius $\theta$ (see, e.g., Peebles
1993). 
In the above equations, $D_L$ and $D_A $ are the luminosity distance
and the angular diameter distance which depend on the redshift $z$ of
the cluster and cosmological parameters: $\Omega_0, \lambda_0$ and
$H_0$.
Using the relation $ D_L = (1+z)^2 D_A$ and paying attention to their
cosmological parameter dependences, the gas mass is written as

\begin{equation}
M_{gas} (<\theta) \propto h^{-5/2} \tilde{D_A}^{5/2}(z;\Omega_0,\lambda_0),
\end{equation}
where $ D_A \equiv h^{-1} \tilde{D_A}$.
If we estimate the gas mass fixing the value of $\Omega_0$ and
$\lambda_0$, it is proportional to $h^{-5/2}$, which is a well
known relation.  


Secondly, we describe the galaxy mass in a cluster.
In general, it is estimated as the total blue (absolute) luminosity
times the mass-to-luminosity ratio (e.g. White et al. 1993).
Note that the total blue luminosity is proportional to $ {D_L}^2 $ and
that the mass-to-luminosity ratio is proportional to $h$ (since it
is derived for nearby galaxies, it does not depend on the
cosmological parameters except for the Hubble constant).
Thus, the galaxy mass $ M_{gal}$ in a cluster is written as

\begin{equation}
M_{gal} \propto h^{-1} \tilde{D_A}^2.
\end{equation}
If we fix the value of $\Omega_0$ and $\lambda_0$, $M_{gal}$ is
proportional to $h^{-1} $.


Finally, we estimate the total mass in a cluster of galaxies using X-ray
data. 
We assume that the intracluster gas is in hydrostatic equilibrium.
Then, the total mass within the radius $R$ is

\begin{equation}
M_{tot} (<R) = - \left. \frac{k_B T_e R}{G \mu m_H} \frac{d \ln
n_e(r)}{d \ln r} \right|_{r=R},   
\end{equation}
thus,

\begin{equation}
M_{tot} (<\theta) \propto h^{-1} \tilde{D_A}.
\end{equation}

\section{Method to estimate cosmological parameters}


The basic assumption of our new method to estimate cosmological
parameters is that the baryon fraction of clusters of galaxies
$M_b/M_{tot}$ is constant.
In this method we do not need to assume that $M_b/M_{tot}$ equals
the global baryon fraction $\Omega_b/\Omega_0$ as in previous works
(e.g., White et al. 1993).
Based on this assumption, we estimate the cosmological parameters.


Using the results mentioned in the previous section,
we can write the baryon fraction ($M_b = M_{gas} + M_{gal}$) in a
cluster as

\begin{equation}
\label{final}
\frac{M_b}{M_{tot}} = A \tilde{D_A} + B h^{-3/2} \tilde{D_A}^{3/2},
\end{equation}
where $A$ and $B$ are some cosmological parameter-independent values.
In this {\it Letter}, we concern with only cosmological parameter
dependence of the baryon fraction, thus precise formula of $A$ and $B$ 
is not needed below. 
In the above equation, the first term comes from the galaxies and the
second from the gas.
As shown in the above equation, the derived baryon fraction depends on
the cosmological parameters through the angular diameter distance. 

The basic and plausible assumption adopted in this {\it Letter} is
that the value of $M_b/M_{tot}$ is constant for all clusters.
However, the derived value of $M_b/M_{tot}$ depends on the adopted
cosmological parameters ($\Omega_0$ and $\lambda_0$) in particular
at high redshift, since it depends on the angular diameter distance
which depends on the adopted cosmological parameters.
That is, if we use another cosmological parameters to derive
$M_b/M_{tot}$, we get another value for it.
If we adopt the correct values as $\Omega_0$ and $\lambda_0$ to
estimate mass of a cluster, we obtain the true value of $M_b/M_{tot}$ and 
it must be constant for all clusters at any redshift.
On the other hand, if we adopt incorrect values for $\Omega_0$ and
$\lambda_0$, we obtain incorrect value of $M_b/M_{tot}$ and it varies
with redshift.
Thus, we can determine $\Omega_0$ and $\lambda_0$ from $M_b/M_{tot}$
as follows: Estimate $M_b/M_{tot}$ for clusters with various redshifts
adopting some fixed values as $\Omega_0$ and $\lambda_0$
(hereafter $M_b/M_{tot}$ vs. $z$ diagram) .
If the derived values of $M_b/M_{tot}$ vary with redshift, then,
adopted values of $\Omega_0$ and $\lambda_0$ are incorrect.
If the derived values of $M_b/M_{tot}$ are constant, then adopted
values  of $\Omega_0$ and $\lambda_0$ are the correct ones.

It is to be noted that derived $M_b/M_{tot}$ depends on $\Omega_0,
\lambda_0$ and $h$ (see eq.(\ref{final})).
However, the Hubble constant dependence cannot be discriminated unless
we know the true value of $B$ in this method.
Although uncertainty of the Hubble constant shifts the absolute value
of derived $M_b/M_{tot}$, it does not change their behavior of the 
redshift evolution. 
Thus, we can determine $\Omega_0$ and $\lambda_0$ without knowing the
value of $h$.
Furthermore, in the above discussion, we took some
assumptions for simplicity, but the final result (eq.(\ref{final})) is 
correct even if we relax some assumptions, for example,
the gas is isothermal. 

In the following, we study $M_{gas}/M_{tot}$ instead of
$M_b/M_{tot}$, for simplicity. 
This is justified when the galaxy mass is negligible compared with the
gas mass ($M_{gal}/M_{gas}$ is about $20 h^{3/2}$ \% in the Coma cluster:
White et al. 1993) and/or the value of $M_{gas}/M_{tot}$ is constant
for all clusters.
In this case, we can use more simple way to estimate $\Omega_0$ and
$\lambda_0$: 
We estimate $M_{gas}/M_{tot}$ using fixed values of $\Omega_0$ and
$\lambda_0$.
We can predict the artificial redshift evolution of the derived
$M_{gas}/M_{tot}$ if the correct values of $\Omega_0$ and $\lambda_0$ are
different from the adopted values.
Thus, we can also determine $\Omega_0$ and $\lambda_0$ from derived
$M_{gas}/M_{tot}$ (using fixed values of $\Omega_0$ and $\lambda_0$)
comparing with expected artificial $z$ evolutions.
In figure 1, we show the expected artificial redshift evolution of
$M_{gas}/M_{tot}$ fixing $(\Omega_0, \lambda_0)=(0,0)$
when the correct values of $(\Omega_0, \lambda_0)$ are 
$(1,0),~ (0,0)$ and $(0.1,0.9)$.
Of course, if the correct values of $(\Omega_0, \lambda_0)$ are
$(0,0)$, the derived values of $M_{gas}/M_{tot}$ are constant.
On the other hand, if $(\Omega_0,\lambda_0)=(1,0) ((0.1,0.9))$,
then derived $M_{gas}/M_{tot}$ increases (decreases) with redshift.
If we can determine $M_{gas}/M_{tot}$ within $\sim \pm 10 \%$ for a
cluster at $z \sim 0.4$ besides for nearby clusters, we can
discriminate three models.

\begin{fv}{1}
{0cm {\begin{minipage}[c]{8cm}
       \psfig{figure=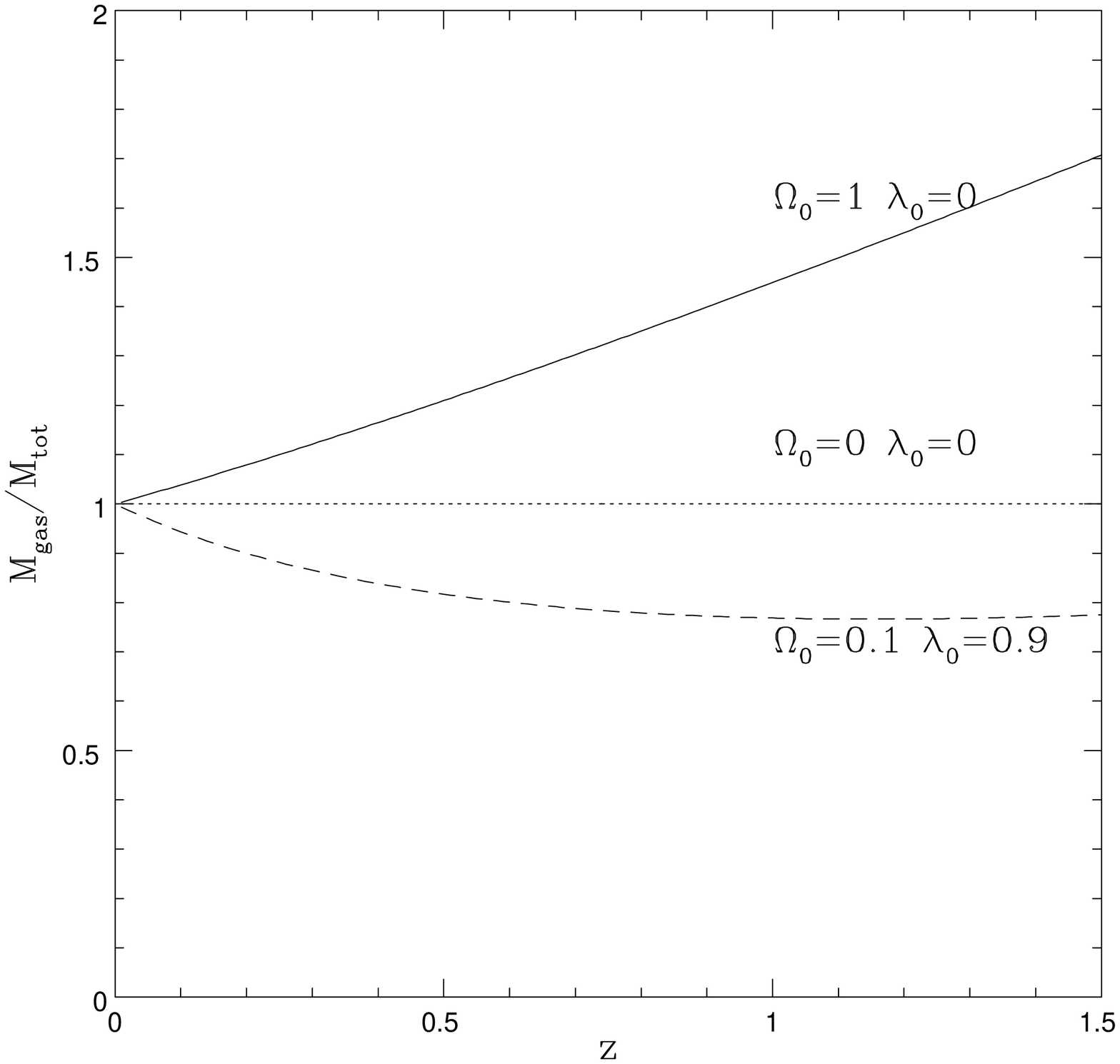,height=8cm}
      \end{minipage}}}
{The expected artificial redshift evolution of the derived
$M_{gas}/M_{tot}$. 
The values of $M_{gas}/M_{tot}$ are estimated fixing $ (\Omega_0,
\lambda_0)=(0,0) $ when the correct values are (1,0) (solid), (0,0)
(dotted) and (0.1,0.9) (dotted).
Normalization of $M_{gas}/M_{tot}$ is arbitrary.}
\end{fv}

\section{Discussion and Conclusions}

As mentioned above, the $M_{gas}/M_{tot}$ (or $ M_b/M_{tot}$) vs. $z$
diagram is a potentially useful method to estimate cosmological
parameters. 
However, to estimate them by this method in practice, we need to check
some uncertainties.
Firstly, the value of $M_{gas}/M_{tot}$ depends on the outer radius of 
the analysis except for $\beta = 2/3$.
Thus, we must know how large a radius is required to derive
$M_{gas}/M_{tot}$.
Secondly, we must know whether $M_{gas}/M_{tot}$ ($M_b/M_{tot}$) is
constant or not, in fact. 
Thirdly, we must know how large a scatter there is.
To check these uncertainties, we need first study $M_{gas}/M_{tot}$
($M_b/M_{tot}$) for nearby clusters which are not affected by adopted
cosmological parameters. 
Unfortunately at the present, we do not have large samples enough to
study this, but in the near future we will be able to have such samples.

At the present, we do not have large samples of
$M_{gas}/M_{tot}$ data enough to study its cosmological parameter
dependences, in  particular at high redshift.
Thus, here, we use inhomogeneous samples which are collected from some 
literatures, to study the possibility of the method mentioned above.
It is to be noted that analysis described below using
inhomogeneous samples is not conclusive, but illustrative at best.
The data are summarized in table 1.
In table 1, we assumed $\Omega_0=0, \lambda_0=0$ and $h=1$.
When the original data are derived using different cosmological
parameters, we translate them using the above relation
(eq.(\ref{final})) to the values for $(\Omega_0, \lambda_0) = (0,0)$
and $h=1$. 
As mentioned above, this data set is inhomogeneous: For example,
outer radius of the analysis (see table 1) and method of mass
estimation are different for each cluster and literature.
In figure 2, we plot $M_{gas}/M_{tot}$ as a function of redshift with
expected artificial $z$ evolution curves. 
In figure 2, the values of data are normalized by the value of the
Coma (which is the leftmost one with error bar).
Since, here, we are interested in only their redshift evolution and
not their absolute values, the normalization is arbitrary.
As is seen, there is a large scatter.
A part of scatter comes from inhomogeneity of the data set.
We think large homogeneous data set will conclude whether this scatter 
is real or not and how large a scatter there is.
If the scatter is real and large, we need some bayon segregation
processes for cluster formation and evolution, for example, energy
input by galactic winds. 
Since we think, in general, there is no efficient process of changing
the baryon fraction averaged within a Mpc scale, we pay attention to
data for which their outer radius is greater than 1Mpc ($\Omega_0=0,
\lambda_0=0$, and $h=1$) and they are shown as large crosses.
The result suggests that high-$\lambda_0$ model seems to be
disfavored. 
However, any conclusion must be left until large homogeneous data set is 
available.

\begin{table}[t]
\small
\begin{center}
Table~1.\hspace{4pt}Data for the baryon fraction.\\
\end{center}
\vspace{6pt}
\begin{tabular*}{\columnwidth}{@{\hspace{\tabcolsep}
\extracolsep{\fill}}p{3pc}cccccc}
\hline\hline\\[-6pt]
cluster & redshift & $ \frac{M_{gas}}{M_{tot}}(<R) (\%) $ & $R (h^{-1} {\rm Mpc}) $ & ref. \\[4pt]\hline\\[-6pt]
Coma        & 0.0232  &  4.95   &   1.5 &  1 \\
A2199       & 0.0299  &  3.2    &   1.2 &  3 \\
A85         & 0.0521  &  6.65   &   0.71 &  2 \\
A3266       & 0.0545  &  5.55   &   0.71 &  2 \\  
A2319       & 0.0559  &  4.31   &   0.70 &  2 \\  
A2256       & 0.0581  &  5.4    &   1.2 &  3 \\
A1795       & 0.0621  &  7.28   &   0.71 &  2 \\  
A644        & 0.0704  &  3.71   &   0.60 &  2 \\
A401        & 0.0748  &  5.0    &   1.2 &  3 \\
A2029       & 0.0768  &  5.7    &   1.2 &  3 \\
A1650       & 0.0840  &  4.17   &   0.55 &  2 \\  
A478        & 0.0881  &  9.05   &   0.98 &  2 \\
A2142       & 0.0899  &  4.99   &   0.97 &  2 \\ \hline 
A3186       & 0.1270  &  6.54   &   0.75 &  2 \\  
A1413       & 0.1427  &  4.07   &   0.86 &  2 \\  
A545        & 0.1530  &  6.40   &   0.91 &  2 \\
A2009       & 0.1530  &  4.81   &   0.65 &  2 \\  
A3888       & 0.1680  &  4.91   &   0.56 &  2 \\  
A2218       & 0.175   &  6.7    &   0.4 &  5 \\
A1689       & 0.1810  &  4.84   &   0.74 &  2 \\  
A665        & 0.1816  &  7.00   &   1.2 &  2 \\
A1763       & 0.1870  &  7.00   &   0.91 &  2 \\ \hline 
A2163       & 0.2030  &  5.94   &   1.1 &  2 \\ 
A1246       & 0.216   &  7.0    &   1.3 &  6 \\ \hline
Cl0016+16   & 0.5545  &  8.44   &   1.7 &  4 \\[4pt]
\hline
\end{tabular*}
\vspace{6pt}\par\noindent
The values of $M_{gas}/M_{tot}$ and $R$ are estimated for $\Omega_0=0,
\lambda_0=0$ and $H_0=100 {\rm km~s^{-1}~Mpc^{-1}}$.
The last column refer to:
1. White et al. (1993).
2. White and Fabian (1995).
3. Buote and Canizares (1996).
4. Neumann and Bohringer (1996).
5. Squires et al. (1996).
6. Kikuchi (1996).
\end{table}

\begin{fv}{2}
{0cm {\begin{minipage}[c]{8cm}
       \psfig{figure=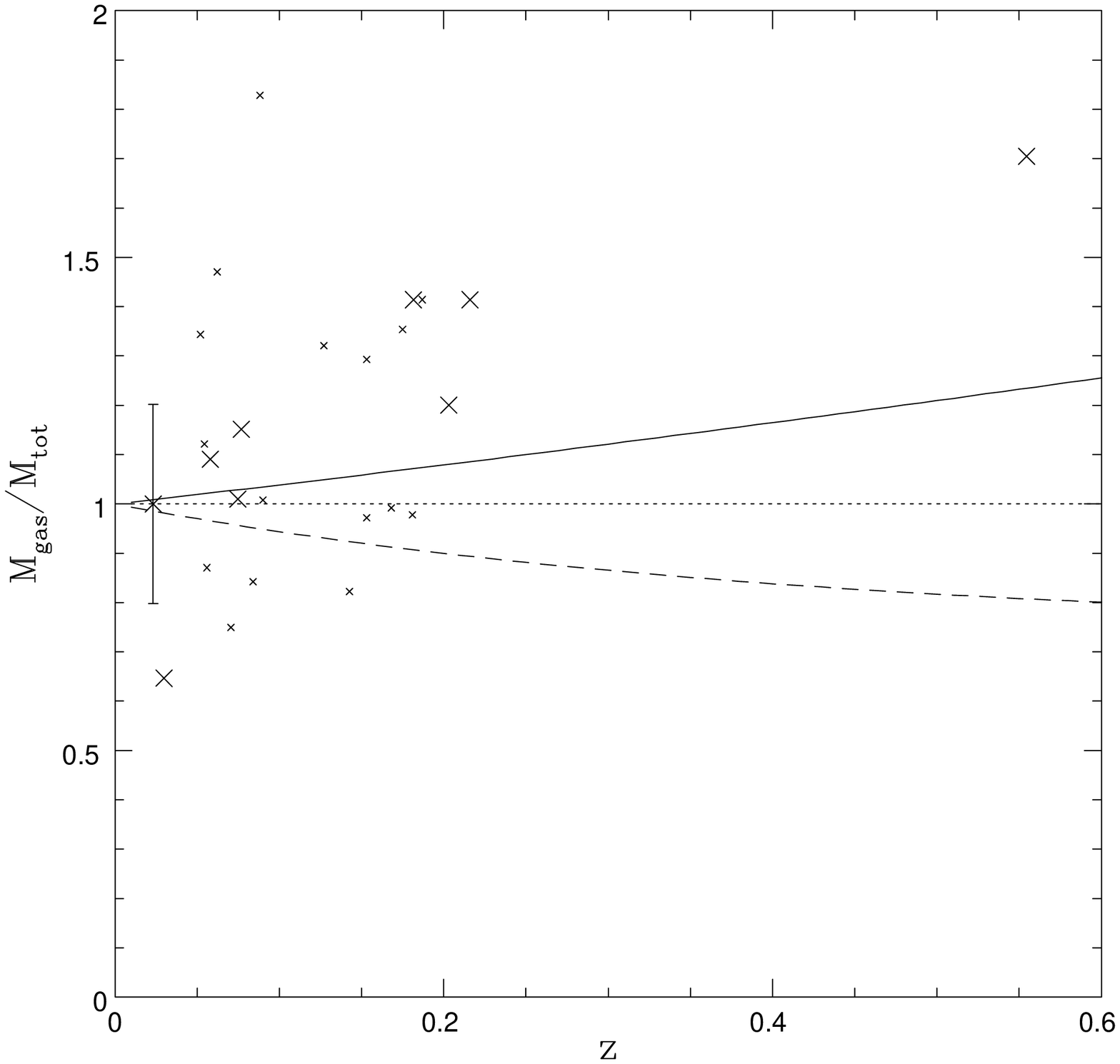,height=8cm}
      \end{minipage}}}
{Same as Fig.1. Crosses are observational data (see table 1)
which are normalized by the value of the Coma (which is the leftmost
one with error bar).
Large crosses indicate clusters with $R>1 h^{-1} {\rm Mpc}$
and small crosses indicate clusters with
$R<1 h^{-1} {\rm Mpc}$.
We assume $\Omega_0=0$ and $\lambda_0=0$.}
\end{fv}

In summary, we proposed a new method to estimate cosmological
parameters using the baryon fraction of clusters of galaxies for a
range of redshifts.
We found that the $M_b/M_{tot}$ vs. $z$ diagram provides information
about the values of cosmological parameters if we can determine the
baryon fraction within $\sim \pm 10 \%$ for a cluster at $z \sim 0.4$.
Furthermore, if the baryon fraction of a cluster is identified with
the global baryon fraction as in many previous works, we can estimate
$\Omega_b$ as $ \Omega_0 \times (M_b / M_{tot}) $ using $\Omega_0$
value derived from the $M_b/M_{tot}$ vs. $z$ diagram.
This is a new $\Omega_b$ estimation method without using big bang
nucleosynthesis studies.  \par
\vspace{1pc}\par
We are grateful to Ken-ichi Kikuchi for providing us with unpublished 
data.
We thank Fumio Takahara for useful comments.

\section*{References}
\re
Buote, D.A., Canizares, C.R.\ 1996, ApJ.\ 457, 565
\re
Kikuchi, K.\ 1996, private communication
\re
Neumann, D.M., Bohringer, H.\ 1996, astro-ph/9607063
\re
Peebles, P.J.E.\ 1993, Principles of Physical Cosmology(Princeton
University Press, New Jersey)
\re
Rybicki, G.B., Lightman, A.P.\ 1979, Radiative Processes in
Astrophysics(Wiley, New York)
\re
Sarazin, C.L.\ 1988, X-ray emission from clusters of
galaxies(Cambridge University Press, Cambridge)
\re
Squires, G. et al.\ 1996, ApJ.\ 461, 572
\re
White, D.A., Fabian, A.C.\ 1995, MNRAS.\ 273, 72
\re
White, S.D.M. et al.\ 1993, Nature 366, 429

\end{document}